\documentclass[manuscript]{acmart}
\usepackage{graphicx}%
\usepackage{multirow}%
\usepackage{amsmath,amsfonts}%
\usepackage[title]{appendix}%
\usepackage{xcolor}%
\usepackage{booktabs}%
\usepackage{listings}%
\usepackage[ruled,vlined]{algorithm2e}

\AtBeginDocument{%
  \providecommand\BibTeX{{%
    \normalfont B\kern-0.5em{\scshape i\kern-0.25em b}\kern-0.8em\TeX}}}
\begin{document}

\title{Complex Event Processing in the Edge: A Combined Optimization Approach for Data and Code Placement}

\author{Halit Uyanık}
\authornote{Both authors contributed equally to this research.}
\email{uyanikhal@itu.edu.tr}
\orcid{0000-0003-2665-2085}
\author{Tolga Ovatman}
\authornotemark[1]
\email{ovatman@itu.edu.tr}
\orcid{0000-0001-5918-3145}
\affiliation{%
  \institution{Istanbul Technical University}
  \city{Istanbul}
  \country{Turkey}
}

\renewcommand{\shortauthors}{Uyanık and Ovatman}

\begin{abstract}
The increasing variety of input data and complexity of tasks that are handled by the devices of internet of things (IoT) environments require solutions that consider the limited hardware and computation power of the edge devices. Complex event processing (CEP), can be given as an example, which involves reading and aggregating data from multiple sources to infer triggering of important events. In this study, we balance the execution costs between different paths of the CEP task graph with a constrained programming optimization approach and improve critical path performance. The proposed approach is implemented as a Python library, allowing small-scale IoT devices to adaptively optimize code and I/O assignments and improve overall latency and throughput. The implemented library abstracts away the communication details and allows virtualization of a shared memory between IoT devices. The results show that optimizing critical path performance increases throughput and reduces delay across multiple devices during CEP operations.
\end{abstract}

\ccsdesc[500]{Computer systems organization~Distributed architectures}
\ccsdesc[500]{Computer systems organization~Sensor networks}
\ccsdesc[500]{Software and its engineering~Software infrastructure}
\ccsdesc[500]{Software and its engineering~Data flow architectures}
\ccsdesc[500]{Software and its engineering~Publish-subscribe / event-based architectures}
\ccsdesc[500]{Software and its engineering~Software performance}

\keywords{Internet of Things, Complex Event Processing, Virtual Shared Memory, Cost Optimization}

\maketitle

\section{Introduction}
As the number of IoT devices increases in smart environments, the need to execute multiple dependent tasks also increases. Various sensor data, such as audio signals, image frames, and temperature values, are used together to make suggestions, raise alarms, check health status, or automate remote tasks \cite{lobaccaro2016review}. Such tasks come with the problem of resource management specifically in edge-IoT devices. One of the paradigms that can be realized in such an environment is complex event processing (CEP). In the context of this paper, CEP is used to represent the process of applying computation, evaluation, or aggregation to data to reach a conclusion \cite{cugola2012processing}. In the context of IoT, a very classical example could be gas leak detection, where gas level data from various sensors can be processed in combination with the images of the area to infer if the gas level readings are consistent with the detected visuals. Other examples of applications are theft detection and health monitoring \cite{9850500, NASERI2022102421, 9900189, computers12110238, KUMAR2024105609}.

A CEP action flow can be composed of multiple parallel event execution sequences that are required to be completed in a limited time frame, where the slowest sequence determines the overall performance of the system. Managing code and data placement becomes a challenge in resource-constrained IoT environments. Scenarios such as danger alarms or health issues require operating with minimum latency, and edge computing becomes essential. The latency issues introduced by the remote cloud server solutions can be overcome by balancing the load between the cloud and the edge computing \cite{ren2019collaborative}. However, cloud options may not be always available, and latency and security requirements might force keeping the computation on edge. Additionally, the cost of cloud solutions can become enormous for individual usage \cite{9145634}. Additionally, the hardware resource requirement of the CEP actions can differ widely from each other, which may prevent the entire flow being deployed into a single device. Different sensors are usually located on different IoT devices, such as a camera and a weight calculator, and therefore CEP solutions should account for this physical restriction on how the raw sensor data should be handled. Hence, code and I/O data may not always be homogeneous, making assignment of code execution location and related hardware resource management critical, and any non-optimal solution can lead to increasing response times \cite{chowdhury2019drls}. Some CEP actions require accessing different data across different IoT devices, which makes a shared memory system a viable solution.

Not all execution requests in CEP are equal in computation cost. While some tasks only query for simple values, such as temperature readings, other steps can include executions that utilize fetching image frames from a data storage and then applying filters on them. When the CEP flow involves a sequence of computationally costly steps (we use the term sequential to indicate that a previous CEP output is utilized as input for another CEP step), it can be hard to assign all code executions to a single device due to hardware resource limitations. Also, due to constant data flow, managing the lifetime of the stored data becomes important to reduce the load on I/O operations. Due to the same reason, the latency should be low so that old data should be processed, if possible, before it is replaced. As the complexity of these tasks increases, the requirement to meet the quality of service (QoS) also becomes a challenge. Although some solutions provide a way to offload some of these tasks to a remote cloud server, this option may not be available due to latency requirements and operational costs. Therefore, computations should be kept on edge if possible while also keeping resource requirements in check.

Latency, in a challenging environment as above, can be decreased with different solutions such as follows. Each individual CEP execution can be migrated between IoT devices depending on resource restrictions. If the code of the CEP is immutable in terms of its rules, the code and its related data can be migrated between devices to reach an optimal throughput for a pre-determined time period. Therefore, the problem is reduced to the management of the distribution of CEP codes among agents on different IoT devices. However, the storage space used to store event data can be limiting. One way to alleviate this issue is to distribute the data related to the specific event code and make it possible for the devices to have access to this distributed data system. To address this issue, a virtual shared memory (VSM) structure can be implemented at the application level. This way, the code execution can access the data in any of the connected devices that expose their stored data. Additionally, the development cost of establishing and managing the communication interfaces of VSM can pose a challenge, and this cost can be reduced with a software library solution to enable faster development. Lastly, while there are multiple ways to optimize code placement and data flow distribution across devices, different solutions come with different trade-offs, making the choice of distribution algorithm an important decision.

In order to deal with the aforementioned challenges, a code and data migration approach is introduced that treats the CEP action flows as a directed acyclic graph (DAG) and optimizes the cost of critical path with the lowest throughput while making sure that the cost of other paths is also as minimum as possible. The proposed architecture also periodically runs the optimization to address the possible CEP critical path change due to dynamic event traffic. Additionally, the proposed approach is implemented in a Python library\footnote{https://github.com/HalitU/cep-vsm-lib} that can be used to configure and distribute CEP flows across IoT devices. The developed library enables creation of CEP flows by registering and configuring I/O event relationship of events that are represented as Python scripts. Then the library can dynamically distribute these individual Python scripts across multiple IoT devices. The proposed library also provides built-in VSM and MQTT communication management. Therefore, the library eases the creation and management of simple CEP flows across small-scale IoT environments. In order to observe the efficiency of the migration optimization, throughput and latency results are compared with some basic intuitive heuristics. 

Contributions presented in this paper can be listed as follows:
\begin{itemize}
    \item Creation of DAG from CEP flows depending on the I/O relationships between events, then optimizing the critical path with lowest throughput and highest latency using constrained programming. Problem definition is detailed in Section \ref{sec:problem_formulation}.
    \item Implementing an open-source CEP engine with VSM that can manage code and data migration and enable remote data I/O across a small number of IoT devices. The CEP engine allows running Python code scripts in addition to queries (temporal conditions, filtering, aggregating, etc.) to enable the execution of more complex tasks. Details of infrastructure and implementation are given in Section \ref{sec:cep_model}.
    \item The proposed optimization approach and heuristics are simulated in an example smart vehicle scenario using raspberry pi devices. Throughput and latency performances are discussed with respect to the critical path problem. The experimental results are provided in Section \ref{sec:experiments}.
\end{itemize}

\section{Literature Review and Background}
This section summarizes previous studies that focus on the use of CEP in IoT environments. Technologies such as VSM, edge computing, code and data management have been previously used to increase the performance of CEP systems.

\subsection{Smart Environments \& Edge Computing} 
Smart IoT architectures such as smart homes and smart vehicles enable automation of different applications such as managing energy consumption rates, raising alarms, and many other critical and non-critical tasks related to daily life. Since the automation process gets more complex as the requirements increase, utilizing cloud based solutions end up introducing latency because migrating a constant event stream to a remote cloud system ends up being too costly. Therefore, edge computing solutions help reduce these problems and enable greater availability of automation systems \cite{s21144932, NASIR2022494, MARTINS202295}. Similarly, smart vehicle systems enable different functionalities varying from managing non-costly but critical engine tasks to covering complex parking and GPS scenarios. The quality requirements of IoT actors also play an important role in making decisions about CEP management \cite{10.1145/3583678.3596884}. Additionally, data collection, aggregation, and shor-term reliability of data is an issue that IoT is confronted with \cite{7380573}. In this study, a smart vehicle case with low latency requirement is covered.

\subsection{Virtual Shared Memory} 
CEP tasks can require different types and sizes of data, such as temperature reading and image frames, across multiple sensors at the same time. This requires managing the latency within I/O operations. One of the ways to address this issue is to utilize a VSM system on application level so that devices can access each others memory or possibly disk locations. VSM is a well studied concept \cite{KARLSSON199779} that has been utilized in a wide range of areas such as machine learning (ML), deep learning \cite{8356235} and malicious attack prevention \cite{8855310}. VSM can be used to access remote data with the cost of additional latency, where the consequent increase in latency degrades the performance of the parallel algorithm \cite{8855310}. In an IoT environment, it may not always be possible to gather all data in one place due to hardware limitations and considerations of energy consumption \cite{7809862}. Therefore, managing the VSM while considering latency poses a challenge.  In this study, VSM is realized on the software layer to enable IoT devices to access each others' data that is stored within an in-memory database.

\subsection{Code Migration} 
There are multiple approaches to migrate executions between IoT devices. If the virtual machine (VM) environments are compatible with each other, it is possible to offload tasks by carrying threads themselves. In \cite{gordon2012comet}, the authors focus on migrating threads depending on their life-time expectancy and possible execution latency. Similarly, \cite{lee2017enhanced} uses the virtual machine environment with the addition of object access latency as a scheduling requirement for code offloading. It is also possible to predict the response time of the CEP operators in advance to place them in their optimal location \cite{cai2018response}. There are studies focusing on QoS and end-to-end latency metrics to optimize operator placement \cite{luthra2021tcep}. In \cite{lindeberg2022study}, tasks are migrated while considering possible missing events in-between the migration process. It is possible to determine operator allocation by looking at how much resource they use, or how scarce their resources become, and consider throughput as a reward mechanism to allocate them \cite{cao2022closed}. In this study, code migration is done by downloading code files from a central IoT device.

\subsection{Data Flow Management} 
I/O storage locations for IoT devices play an important role in CEP latency performance. In the context of this study, data flow management is the process of managing the I/O storage locations of data related to CEP events. In \cite{li2021queec}, the authors propose a framework in which the required task library can be downloaded to IoT devices using a cloud platform, and then throughout the execution cycle, these binaries and corresponding execution data can be transferred to other IoT devices with minimum latency. Another approach is to replicate the data \cite{salah2022adaptive, mohamed2023aoeho}, which enables the execution of multiple queries on different devices on the same data. However, it should be noted that some IoT devices are limited in memory and storage capacity. In \cite{luthra2019inetcep}, the authors detail the limitations caused by only the consumer or the publisher initiating the communication, and propose an improvement to the content centric network architecture that allows both of these elements to run together to reduce communication delay. Another study focuses on extending the Apache Storms scheduler interface to consider worker queue sizes to improve data incoming routing, as well as introducing failover mechanisms \cite{sun2020dynamic}. Another issue is to handle the overload caused by continuous events; in such cases, it becomes important to consider which events to discard and which events to assign higher priority \cite{9309088}. In \cite{9626151}, the authors evaluate the approach based on the bus and the flow of enterprise services using CEP tasks. There are also studies that combine data and code placement and decide assignments with heuristic solutions to the generalized assignment problem \cite{8271954}. This study manages data flow by managing where the event results are written and migrating already existing data between devices whenever the target storage device changes.

\subsection{Other Related Studies} 
In addition to the studies mentioned above, there are approaches that focus on using machine learning (ML), deep learning (DL), and reinforcement learning (RL) to learn how the system reacts throughout a certain period of time and then manage the distribution of tasks and data accordingly \cite{verma2022fetch, zhang2022task, huang2022reinforcement}. Some studies focus on dealing with heterogeneous environments and corresponding vendor conflicts by providing new pipeline architectures \cite{dehury2022toscadata}. This study experiments on homogeneous RPi4B devices, but the developed CEP engine can be used on any device that supports the required Python libraries. Dynamic network conditions, such as disconnects, are handled for workers, as the optimization distributes the work to currently available workers at periodic evaluation intervals. There is a single point of failure for the device that runs the CEP optimization.

There are studies that allow edge devices to share GPU power, but these are usually devices with higher hardware capabilities \cite{CECILIA202314}. For cases such as image processing tasks that utilize GPU tools and large volumes of data in edge IoT devices, solutions should cover cases where non-relevant data can be discarded to reduce disk usage \cite{DELUCIA2023207}. Some studies focus on managing the high load on edge computation, such as video events, and improving overall latency and throughput \cite{9411841}. Instead of focusing on small locations such as factories and homes, some studies focus on smart cities and the way data are managed \cite{9018282}. Another study similar to this paper uses the DAG relationship to consider grouping of actions to be executed placed on the same device to prevent early consumption \cite{8594546}. The authors apply a grouping process on the one-hop difference between the tasks, unlike this study which considers the dependency relationship on all of the tasks on a critical path. The authors in \cite{9457745} consider a constrained problem that focuses on resource consumption and delay costs for a time-dependent relationship between the detection times and the actual point of the problem occurring. In this study, complex tasks such as the mentioned image processing are covered, but with limited functionalities. Additionally, this study includes data and code placement locations, relationship between sequential events, migration costs due to device changes, remote I/O penalties, number of events that a device can run in the constrained programming optimization in order to have a wider coverage of problem parameters. The scaling of IoT devices and future challenges are also well studied in the literature \cite{ATZORI20102787, 10.1145/3301443}. 
Further studies focus on robotic IoT, including more complex computation requirements and challenges faced while executing ML models \cite{10.1145/3663338.3665828}.

\section{Problem Formulation} \label{sec:problem_formulation}
This section explains how the events of the CEP and corresponding I/O topic subscriptions/publications can be represented in a graph relationship. This graph is then extended by including the raw data producing applications, so that the data output location of these applications can also be included in the optimization cost. The critical path of the graph is identified as the path where the throughput is lowest, and this path is aimed to be optimized so that the minimum throughput is increased.

\begin{figure}[ht!]
    \centering
    \includegraphics[width=0.7\linewidth]{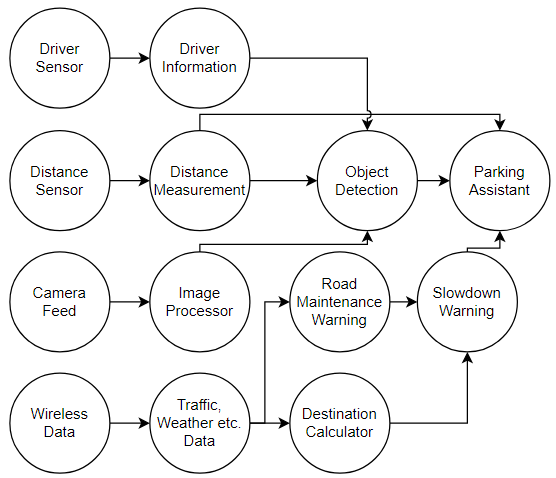}
    \caption{Example dependency relationship between CEP tasks for the smart vehicle scenario.}
    \Description{Example dependency relationship between CEP tasks for the smart vehicle scenario.}
    \label{iot_case_cep}
\end{figure}

\subsection{Critical Path Problem} 
In a scenario where there are no cyclic dependencies between the dependent CEP tasks, the connection between them can be represented in a DAG (see Figure \ref{iot_case_cep}). When tasks are distributed randomly, it is possible that some executions in the flow may be executed in longer times due to resource limitations and I/O delay. Furthermore, the latency and throughput of the system can be represented within the edges between different code steps, indicating how many decision making executions are made in a time period. Then, the critical path of the system can be represented as the path which has the highest cost. The cost function is determined according to latency and throughput. The goal of this study is to implement an optimization that can decide on the code and data location assignments and balance the path cost across all paths of a CEP flow.

\subsection{Optimization Goal} 
The main aim of optimization is to increase the throughput of the critical path within the CEP flow. For this purpose, different metrics are collected that describe the flow of data, the location of stored data, and the location where a task is executed. Optimization aims to reduce the negative effects of changes in data traffic and minimize cost. To achieve this, the code as well as the event data locations are assigned so that the throughput and delay is optimized within a time frame. Whereas the goal of optimization is to stabilize the amount of changes while reaching maximum throughput within the system, the system load, parameters, flow, and data types can always change. Therefore, it is important to indicate exactly what the optimization goal covers. The optimization is run periodically every 30 seconds. After each evaluation, the code and data location updates are published to the worker devices. Optimization aims to improve average throughput and minimize average latency compared to heuristic approaches, which will be discussed in the experimental section.

\subsection{Mathematical Background}

This section details the mathematical model of the approach that covers the critical path as well as the representation of the CEP as a graph. Then it presents the cost calculation and the optimization model. Mathematical notations are given in Table \ref{table:Notations}. Throughout its lifetime, proposed architecture collects different statistics such as data I/O latency, task execution latency, and byte size of data processed to optimize and adapt to the current state of the IoT environment.

\begin{table}[ht!]
\caption{Notations}
\label{table:Notations}
\begin{tabular}{ll}
\toprule
Notation & Definition \\ 
\midrule
$F$ & A CEP flow composed of steps \\
$S_i$  & A CEP step consisting topics and an action \\
$E_i$  & Subscribed topics for a step \\
$A_i$  & Action to be executed at step \\
$P_i$  & Path consisting one or more steps \\
$W_i$     &    Worker devices        \\
$N$ & Number of worker devices in the system \\
$loc_{A_i}$ & Execution locations for an action of step \\
$loc_{D_i}$ & Location of the data for step to be stored at \\
$lat_{P_i}$    & Latency of a path \\
$lat_{S_i}$    & Latency of a step \\
$lat_{read}$    & Latency of a reading input data \\
$lat_{execute}$    & Latency of executing an action \\
$lat_{write}$    & Latency of writing output \\
$Act_S$    & Activation cost of a step \\
$lat_{download}$    & Time it takes to download code files \\
$lat_{subscription}$    & Time it takes to subscribe to topics \\
$data_{bytes}$ & Size of data processed by event \\
$Cost_{S_i}$ & Cost of a step \\
$Cost_{P_i}$ & Cost of a path \\
\bottomrule
\end{tabular}
\end{table}

A flow step ($S_i$) is identified as a tuple relationship that is composed of an event input ($e_i$) that triggers the action, the corresponding output ($o_i$) produced from it, and the execution of the triggered code ($a_i$). Therefore, an output can also be the trigger input of another step. A flow can consist any number of these steps which may or may not run in parallel. System is also constrained by the fact that no cyclic relationships can exists between these steps. Therefore, collection of these flow steps composes a DAG. A path ($P_i$) in the DAG is identified as a set of steps which starts from the source nodes of the DAG until its sink nodes. A DAG can therefore also be identified as the collection of paths as well. Mathematical representation of steps, paths and related CEP components are given in Figure \ref{cep_mathematical_definition}.

\begin{figure}
\[ F = \{E \cup A, \succ_c \} \]
\[ E = \{ e_0, e_1, ... e_n \} \]
\[ A = \{ a_0, a_1, ... a_n \} \]
\[ S = \{s_0, s_1, ...,s_n\}  \]
\[ s_i \in S = \{(e_i, a_i, o_i) | e_i , o_i \in E , a_i \in A \} \]
\[ \succ_c = \{ (s_i, s_j) | S_i \in S \quad and \quad S_j \in S \} \]
\[ s_{0} : \forall i > 0 \ \forall c \ge 0 \ s_{0} \nsucc_c s_{i} \]
\[ s_{i} : \forall j > i \ \forall c \ge 0 \ s_{j} \nsucc_c s_{i} \]
\[ P = \{p_0, p_1, ...,p_n\}  \]
\[ p_i \in P = \{(s_i...s_{i+j}) | 0 \le i \le j \le n \quad and \quad s_i \in S\} \]
\caption{Path and step definitions that compose the CEP.}
\Description{Path and step definitions that compose the CEP.}
\label{cep_mathematical_definition} 
\end{figure}

Each action (code) can be active or passive in a worker, and the data for an action can be stored in one of the workers. This introduces the first constraint (Equation \ref{constraint1}) of the problem, which is the necessity of distributing a task to at least and at most one device. Similarly, each data output should be written to only one of the devices (Equation \ref{constraint3}). The reason and assumption here is that since IoT devices usually have less hardware capabilities, each code can be assigned to only one device to prevent overload on resource consumption.

\begin{displaymath} 
loc_{action} = m_{ij} \in \{0, 1\}^{|W|\times|A|}
\end{displaymath}
\begin{displaymath}
loc_{data} = l_{ij} \in \{0, 1\}^{|W|\times|A|}
\end{displaymath}

\begin{equation} \label{constraint1}
\forall a_i \in A \sum_{j=1}^{j=W} m_{ij} = 1
\end{equation}

\begin{equation} \label{constraint3}
\forall l_i \in loc_{data} \sum_{j=1}^{j=W} m_{ij} = 1
\end{equation}

The latency of a step can be identified as the amount of time each event takes from the time it should be triggered until the time it produces the output event. This consists of the time to read the data, the time to execute the action and the time to publish the output data to VSM (Equation \ref{eq:step_latency}). The latency of a path can then be identified as the sum of the step latency that composes it (Equation \ref{eq:path_latency}). The latency of the critical path is the path with the highest accumulated latency. The end goal is to minimize the maximum of these latency values across all CEP DAG paths. Another description could be that the critical latency is the maximum delay of the consumed event that was produced by a sensor that arrived last in the last code task within the CEP flow, as this delay includes the execution and data read/write times across that specific path steps.

\begin{equation} \label{eq:step_latency}
lat_{S_i} = lat_{read} + lat_{execute} + lat_{write}
\end{equation}
\begin{equation} \label{eq:path_latency}
lat_{P_i} = \sum_{i=1}^{j}{lat_{S_i}}
\end{equation}

Throughout the lifetime of applications, each step activation also has its time cost which affects the overall performance of the system, whereas this cost is not utilized in the optimization for this study, it is an important factor that needs to be mentioned. The activation cost is the sum of time it takes to download a code script and subscribe to the event triggering topic (Equation \ref{migration_cost}).

\begin{equation} \label{migration_cost}
Act_{S_i} = lat_{download} + lat_{subscription}
\end{equation}

When calculating the data fetching $lat_{read}$, result writing $lat_{write}$ and task execution $lat_{execute}$ latency of the steps on a path, different costs can be incurred depending on data location. If the previous step $s_{i-1}$ writes its output $o_{i-1}$ to the current device $W_i$, then it will be easier to retrieve the data, and it will be harder otherwise (Equation \ref{eq:read_penalty}, \ref{eq:write_penalty}). These two relations can be used in optimization with the aforementioned constraints to minimize the overall path latency. Therefore, the latency of a step can be further detailed with custom penalty multipliers.

\begin{equation} \label{eq:read_penalty}
    lat_{read_i}=
    \begin{cases}
        T_{read}              , & \text{if } i==W_i \\
        \alpha \times T_{read}, & \text{otherwise}
    \end{cases}
\end{equation}

\begin{equation} \label{eq:write_penalty}
    lat_{write_i}=
    \begin{cases}
        T_{write}              , & \text{if } i==W_i \\
        \beta \times T_{write}, & \text{otherwise}
    \end{cases}
\end{equation}

In addition to latency, throughout a time period, the data size that is processed across the paths are also important, as it is possible that latency can be very small over a very cheap event that does not consume much resources, but another event that processes images could run very slow due to non-optimal distribution. Therefore, the cost of a step is identified as a combination of time and data size (Equation \ref{eq:step_cost}), and optimization is focused on minimizing the sum of step costs over a path (Equation \ref{eq:path_cost}), for all paths in the CEP diagram (Equation \ref{optimization_goal_func}). 

\begin{equation} \label{eq:step_cost}
Cost_{S_i} = \frac{lat_{P_i}}{data_{bytes}}
\end{equation}

\begin{equation} \label{eq:path_cost}
Cost_{P_i} = \sum_{i=1}^{i=N} Cost_{S_i}
\end{equation}

\begin{equation} \label{optimization_goal_func}
\forall i \in P \left(min(Cost_{P_i})\right)
\end{equation}

\subsubsection*{Complexity of the optimization} The initial generation of the paths for the flow graph from raw sources (R) and actions (A) takes $O(R + A*p)$. Where p is a static number for each event that defines the sum of the number of topics consumed and published that is used to create edge relationships in the graph. Then the statistics of the execution of events from the workers (W) are processed, which introduces a cost of $O(W * A)$. The objective function is prepared by visiting each path of the flow graph, where the number of distinct paths is $O((R + A) + E)$, where (R + A) is the number of nodes and E is the number of edges that connect these nodes. Lastly, the CP-SAT optimization solver \cite{cpsatlp} is used to solve the constraint programming problem.

\section{IoT CEP Management System Model} \label{sec:cep_model}
This section first describes how the IoT CEP infrastructure is established. The CEP code migration and data flow management process is then explained. Third, an example real-world scenario that is taken as a basis for experiments is depicted in detail. Lastly, a brief explanation on CEP distribution algorithms is provided.

\subsection{IoT - CEP Infrastructure}
\begin{figure}[ht]
    \centering
    \includegraphics[width=0.6\linewidth]{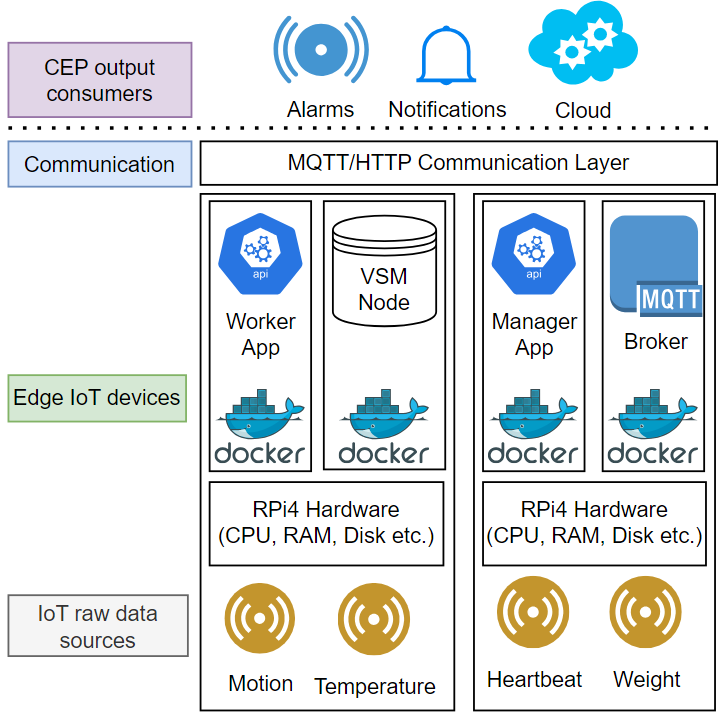}
    \caption{IoT edge infrastructure with CEP worker and manager applications.}
    \Description{IoT edge infrastructure with CEP worker and manager applications.}
    \label{system_architecture}
\end{figure}

The system architecture for the experiments and simulation environment can be seen in Figure \ref{system_architecture}. A device that stores the MQTT \footnote{https://mosquitto.org/} message broker acts as the management unit. The CEP application in this device is responsible for collecting statistical data from workers and determining how the CEP codes and data flows will be distributed. The other type of device is the worker, which is made up of a CEP application and the MongoDB database unit. In Linux machines, the temporary file system (tmpfs) can be used to store data in memory, which is used in this study to store the Docker MongoDB container data to create a virtual memory space. Whereas tmpfs comes with the disadvantage of losing data on device restart, it also enables faster access to data. Considering the fact that the lifetime of the data in the study scenario is short, tmpfs can be utilized for faster event execution with the probability of losing very short-term data. Another benefit of choosing MongoDB is that it provides an easy way to code querying/aggregating data operations, which is desired in CEP systems. Different MongoDB instances across the IoT environment can be connected by any worker device through the software layer which enables the VSM system. In addition to the message broker, CEP applications also utilize HTTP/REST requests to send heartbeats between each other, download CEP codes, or collect execution statistics. Excluding the initial steps to install the required software, all communication is constrained within the distributed IoT devices, and no other external connection is required.

The developed library has multiple components that enable constant management of the CEP flow. The management module enables event streams as mentioned earlier and manages the code and data assignment distribution across workers. The data module manages the VSM and the connections between the sensors. The last module enables downloading the CEP code and starting the consumer/publisher processes on workers to actually start CEP flow working. Supported CEP codes can either be executed on incoming raw sensor data directly, such as audio signals or temperature values, or they can be executed on queries that can read a single or series of data from one or multiple data locations within requested time periods. The execution results are then published on the requested broker topics. Distributions are done periodically for connected devices at the evaluation time; therefore, if any new IoT worker is connected, its resources will be taken into account without tampering with the management device. Lastly, the workers know the IP of the management device which is given within a configuration file initially.

\subsubsection{Initialization of Distributed IoT CEP System} 
\begin{figure}[ht]
    \centering
    \includegraphics[width=0.6\linewidth]{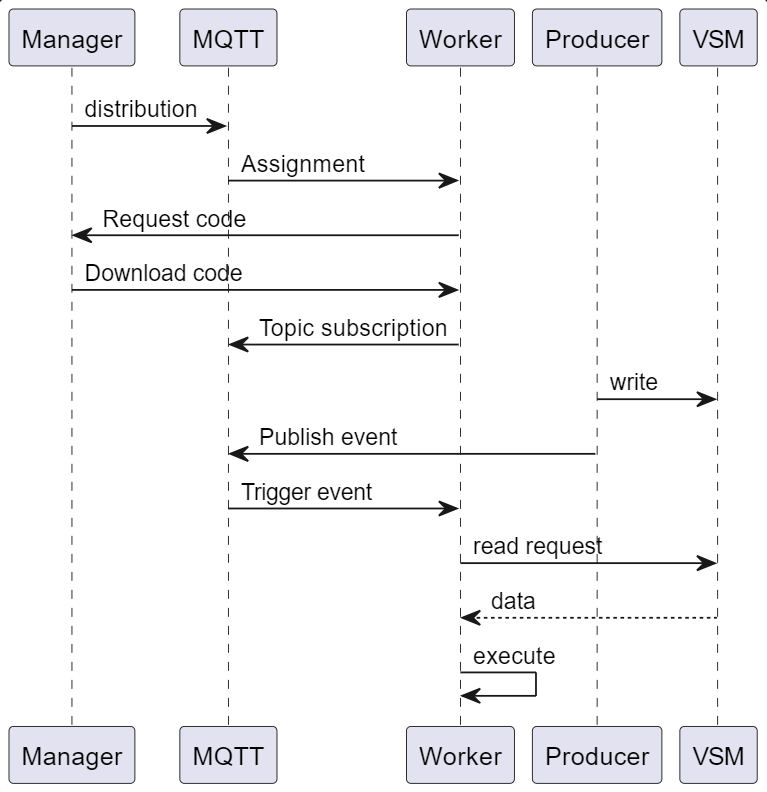}
    \caption{CEP steps between distribution and code execution.}
    \Description{CEP steps between distribution and code execution.}
    \label{startup_flow}
\end{figure}

Figure \ref{startup_flow} shows the basic steps taken between the time that a manager device distributes assignments and the time that a worker device executes an event. When applications first start up, specific steps are followed to initialize the connection between the CEP system devices. Considering the fact that the applications can run concurrently, this is the "happy path" initialization sequence;

\begin{itemize}
    \item The application, database, and MQTT containers start running. The HTTP endpoints that provide heartbeat and code files are exposed.
    \item The server waits until it can connect to the message broker and initialize the management topics.
    \item In the meantime, clients periodically check the server to see if it is ready to manage the cluster.
    \item When clients receive a ready response, they create their own topic consumers for management topics.
    \item The client then publishes their database host and port information to other devices. When a client receives such a message, it redistributes its own info in-case the source client has not received it yet due to starting up late.
    \item After publishing the database information, the client starts initiating the required topics for the events if they are not already created yet, then creates corresponding consumers with a trigger to assigned configuration event. At this point, the system is up and running.
\end{itemize}

Devices within the distributed environment can access each others' in-memory database which contains the information required throughout the CEP flow. This is enabled with an exposed interface on the software layer that provides access to the VSM environment. Devices announce their database information to other workers through the management device whenever they connect to the cluster. Workers use database connections to retrieve input event data from whichever device it is stored on.

\subsection{Code Migration \& Data Flow Management}
In order to manage the CEP executions, the management device periodically monitors the performance of the system. The algorithm \ref{cep_dist_algo} summarizes how the critical path optimization is applied. Since the initial distribution cannot utilize any execution statistics, it uses a round-robin approach to distribute the tasks. After the first period is over, the management device starts collecting execution statistics. Then it runs the function of Equation \ref{optimization_goal_func} to figure out optimal assignments on all paths. The management unit then sends the code and data assignment messages to the workers to carry out the CEP tasks. When a worker receives a code management message, it first checks if it is an activation or deactivation request. If former, it starts the code migration process which requests to download the code files from the management device, and after downloading and importing its module, it subscribes to event topics for consumption. At this point, the system can start processing events. If the message was a deactivation request, then the worker unsubscribes from the topics and deletes the downloaded code file. The message sent by the management device also includes the designated location of the event data. Lastly, if any data are still present in the old VSM location, the newly assigned device migrates these to their new location.

\begin{algorithm}[ht]
    \KwData{CEP \& worker settings.}
    \KwResult{Code and data I/O assignments.}
    Collect execution statistics\;
    Create DAG via pub/sub topics of CEP\;
    \For{each path $P_i$ within the flow $F$}
    {
        Calculate sum of costs of all steps $s_i$\;
    }
    Run function in Equation \ref{optimization_goal_func}\;
    Optimization provides assignments such that the cost from head to sink in the flow $F$ is nearly equal for all paths as much as possible\;
    \textbf{return} $assignment$\;
    \caption{Code \& data I/O assignment management using constrained programming}
    \label{cep_dist_algo}
\end{algorithm}

\subsection{Code Registration} 
Registering a CEP flow to the library requires several steps to follow. The first step is to set up a configuration file that consists of the necessary information for the library to establish communication in a distributed manner. Once the applications are running, the management unit handles all the distribution processes. Therefore, the initial registration of the real-world CEP flows is done by coding the corresponding flow details using the library. The library handles the DAG creation process by using the input/output topic information given in these initial configurations.

\begin{figure}[ht]
    \centering
    \includegraphics[scale=0.55]{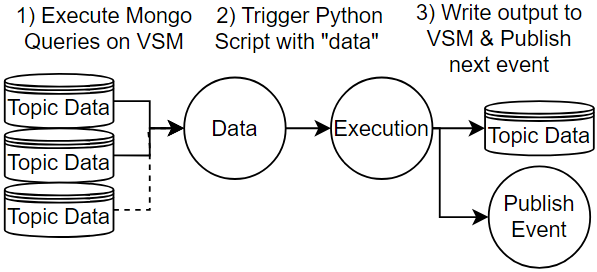}
    \caption{CEP data collection and script execution flow.}
    \Description{CEP data collection and script execution flow.}
    \label{vsmToDataSample}
\end{figure}

The library requires several configuration parameters to register the CEP steps (a minimal setup is shown in Table \ref{configurationParameters}). Briefly, the library requires where to read the code file (action\_path), to which topics it should subscribe to (required\_sub\_tasks, output\_topics) and publish to, as well as optional arguments (optional\_args) and query that enables fetching the data from the VSM (query). The \textit{query} parameter is executed on the location of the data (VSM) of topics registered by an event step, and the gathered data is used to trigger the Python script (as shown in Figure \ref{vsmToDataSample}). The required event data are fetched individually, then aggregated into a single object on application level before triggering the Python script. The \textit{query} parameter is used to trigger the following CEP operations:

\begin{itemize}
    \item Select specific data columns from VSM.
    \item Apply conditional querying to filter according to specific numeric or boolean values, or limit the query to specific time window, etc.
    \item Order the fetched data according to a field.
    \item Limit the number of rows taken.
\end{itemize}

\begin{table}
\caption{Configuration Parameters Used to Register Code Scripts to CEP Library}
\label{configurationParameters}
\begin{tabular}{ll}
\toprule
Field & Definition \\ 
\midrule
action\_name  & Name of the python method to execute  \\
action\_path  & Code location in file system  \\
required\_sub\_tasks  & Topics to subscribe to \\
output\_topics  & Topics to publish execution info \\
optional\_args  & Optional arguments for the code \\
query & Mongo query to execute in VSM \\
\bottomrule
\end{tabular}
\end{table}

After the event is registered, the code can be deployed considering the message broker and other necessary infrastructure is ready. Unlike most of the previous studies that introduce new syntax for efficiently coding CEP codes, this study uses an existing programming language (we use Python in our study). This choice also supports additional method execution capabilities not limited to aggregation, window-fetching, or value mapping. These can be left to developers to decide, the developed library only provides means of connecting the VSM data to the required script to be executed and the respective data and script distribution.

The CEP engine considers each Python script as an atomic unit and distributes these scripts without decomposing them into smaller pieces. In the experimented scenario various different types of Python tasks are used varying from simple conditional checks to euclidean distance calculations, and from Gaussian filters to object detection.

\subsection{Smart Automotive Scenario}
To evaluate the library and the CP optimization approach, a smart vehicle scenario is considered, the CEP flow of which is shown in Figure \ref{iot_case_cep}. Within the flow, there exist sensors that continuously produce raw data that trigger the intermediate CEP events. These mainly produce two different data types, size-wise small values that can be evaluated in mass, such as distance measurements, and comparably larger data that are processed one at a time, such as camera footage(image frames). Similarly to real-life, during the time a car travels across a road and tries to find a parking spot, then attempts to park to that slot, different events may occur. Such as possible sudden obstacles, which can be small animals or walls, and other physical objects. The intermediate events are composed of different functions, such as object detection, which utilizes an already trained model working on image frame, or basic mathematical calculations for distance measurement or boolean operations. 

Due to the nature of vehicle movement, events need to be executed in a short duration. Additionally, the CPU and RAM costs for these events are not equal. Throughout the experiments, similar to real-life, the size of the data provided by the data producing sensors changes, which changes the overall cost of the steps, and therefore the critical path itself. This creates a constant optimization demand for the code and I/O locations to reduce possible throughput loss. Lastly, the experiment results focus on the performance of the last event, at which all CEP paths end. As the throughput and delay of the critical path improve, the overall throughput of this last event will also improve. In our experimental setup, there are 9 sensors that act as worker units, which also run the raw sensor producing applications. There is a tenth device that acts as the central management unit as explained in the previous sections. All devices run continuously, and CEP distribution evaluations are done in 30 second periods.

\subsection{CEP Distribution Algorithms} 
In this study, the goal of the distribution algorithms is to utilize CEP DAG and sensor event execution statistics, so that the management device can optimize code and I/O placement locations. Since the placement of data affects the I/O delay, due to limited hardware resources, each of the CEP event codes and related I/O can be located at a single device. Additionally, the constrained programming approach utilizes a basic depth-first search to find all simple paths in a DAG. At each evaluation period, performance and resource statistics are collected from the sensors. Then the management device combines the collected information to evaluate the previous distribution and decide on the next code and I/O locations. Note that these steps are not mandatory for all heuristic approaches. Distribution approaches, such as randomizing execution locations, do not need to check the graph structure of the CEP flow or the recent execution statistics from the worker devices. Therefore, inherently simple heuristics can have distribution times reduced than those of the constrained programming approach.

As mentioned in the literature section, there are many different solutions that can be applied to task distribution problems. However, specifically in this CEP distribution scenario, a distribution needs to be made in short times to prevent possible incidents, and require low size of data to run since the hardware resources are limited. Therefore, in this study, the execution time of the CP is limited to 10 seconds. It is observed that during the experiments 2.5 seconds on average was enough to find an optimal solution. Another issue with the distribution is to determine in what periods the algorithms will run. Keeping a short window between two distribution runs can impact the performance of the overall system due to possible constant migration costs. This also shows one limitation of the system, if the event stream patterns are too chaotic, for example, if the volume and variety changes each second, then the devices may not be able to keep up an optimal distribution all the time. However, keeping the distribution window too long makes it possible to miss optimal assignments for any potential short-term data traffic and consequent execution statistics that require attention.

\section{Experiments} \label{sec:experiments}
The experiment environment consists of 10 Raspberry Pi4 (RPi4) \footnote{https://www.raspberrypi.com/} devices. Each device has 4GB RAM, 16GB disk space, a single processor, continuous power supply, and cabled internet connection within a virtual private network (VPN). It is also assumed that devices run without hardware failures throughout the experiments, and fault recovery is not considered within the scope of this study. However, since the hardware capabilities of the devices can be considered too powerful to simulate constrained IoT scenarios, in one experiment configuration, device CPU usage is limited to 0.5 using Docker and underlying operating system configurations. Experiments are done through a set of 30 minute periods in which the management device decides on the next distribution every 30 seconds. This time is chosen intuitively after considering the distribution window problems mentioned in the previous section. The CEP action statistics consisting of the total number of event executions, data read/write times, data delays, and some performance statistics of the paths are collected from the worker sensors by the management device at the start of each evaluation period. Additional metrics such as code activation times and distribution times are also collected for a more detailed evaluation. During the simulation, it is assumed that no network or hardware problems occur.

\subsection{Evaluation}
The Constrained programming optimization approach, as well as three heuristic distribution approaches, are tested within the developed library. Since the bottleneck in any flow path shows its negative effect on the performance of the last event, in order to evaluate the performance of the critical path, the performance of the last event is reported. Therefore, it should be noted that these results are not an indicator of the maximum possible performance of the intermediate events. The summary of the experimented distribution approaches are;

\begin{itemize}
    \item Complete round-robin (CRRB): Worker devices are sorted without any performance considerations, and CEP tasks are assigned to these devices one by one in round-robin format.
    \item Random device (RANDOM): CEP tasks are distributed using a uniform random distribution across connected devices. The task count per device is balanced to prevent unnecessary load on the same devices. Distribution is also done once at the start of the simulation and kept as is throughout the simulation to prevent continuous migration overhead. This approach is repeated 25 times in each scenario to observe the standard deviations.
    \item Locality (LOCAL): Each composite event is greedily placed on the device where the stored event input data size is highest. The remaining input data write location(s) is also changed to this device. The code capacity of each worker is limited to two to prevent resource scarcity. As a side note, this approach fails most when the data production rate oscillates frequently, due to possible continuous migration and forwarding costs.
    \item Genetic Algorithm (GA): GA works by first generating an initial population, whose individuals are feasible solutions to the problem. Each individual represents the solution to the CEP code and data distribution problem by encoding the event output locations and execution locations in an array format for all steps in all paths. Next, GA generates several generations using select, cross, and mutation operations. This loop allows for the generation of better performers \cite{10.1145/3140256, 10.1007/978-3-031-22677-9_32}. However, one downside is that the operation depends on randomness because keeping mutations and cross operations deterministic may keep the solution at a local maxima. Another issue is that GA may require running several generations to reach a satisfactory feasible solution. However, since the case study covers a car scenario where optimization speed is important, GA performance may not perform well under restricted parameters. Therefore, the GA parameters are kept as follows, population size is set to 200, generation count is kept to 20, each generation keeps top 5 individuals to keep best performing feasible solutions, after the remaining 195 individuals are generated from the initial population using cross operation, quarter of the population is mutated with a chance of 0.5. This loop continues until the generation count is reached. During our experiments with the CPU restricted devices, GA runs around 10 seconds to process all generations. The fitness function for the GA returns the inverse of the maximum cost on all paths. Additionally, device limitations that CP uses are also applied to GA to keep the fitness of invalid and poor performing individuals at 0. 
    \item Constrained programming (CP): All events are placed according to the solution obtained from the constrained programming approach mentioned in the previous section. Additionally, in order to introduce the effects of migration cost and corresponding message loss when the execution device of a task changes, CP objective function is multiplied with a device change penalty multiplier. 5 different penalty multipliers, which are chosen intuitively, are tested. These are 1.0, 1.25, 1.5, 1.75 and 2.0. 1.0 indicates that there is no penalty and optimization can change target devices as long as the distribution satisfies other requirements. Penalty multipliers can be used to adjust the optimization behavior considering the constraints on the migration capabilities present in the IoT environment under study.
\end{itemize}

There are three different configurations for each of the performance metrics. First, the CP configurations are tested between each other (a), then one of the CP configuration is tested with the heuristics (b), lastly, the CPU capability of the devices are halved to observe if results stay consistent between different device constraints (c). Each distribution is run at least 25 times to observe deviations.

\begin{figure*}[ht!]
    \centering
    \includegraphics[width=1.0\linewidth]{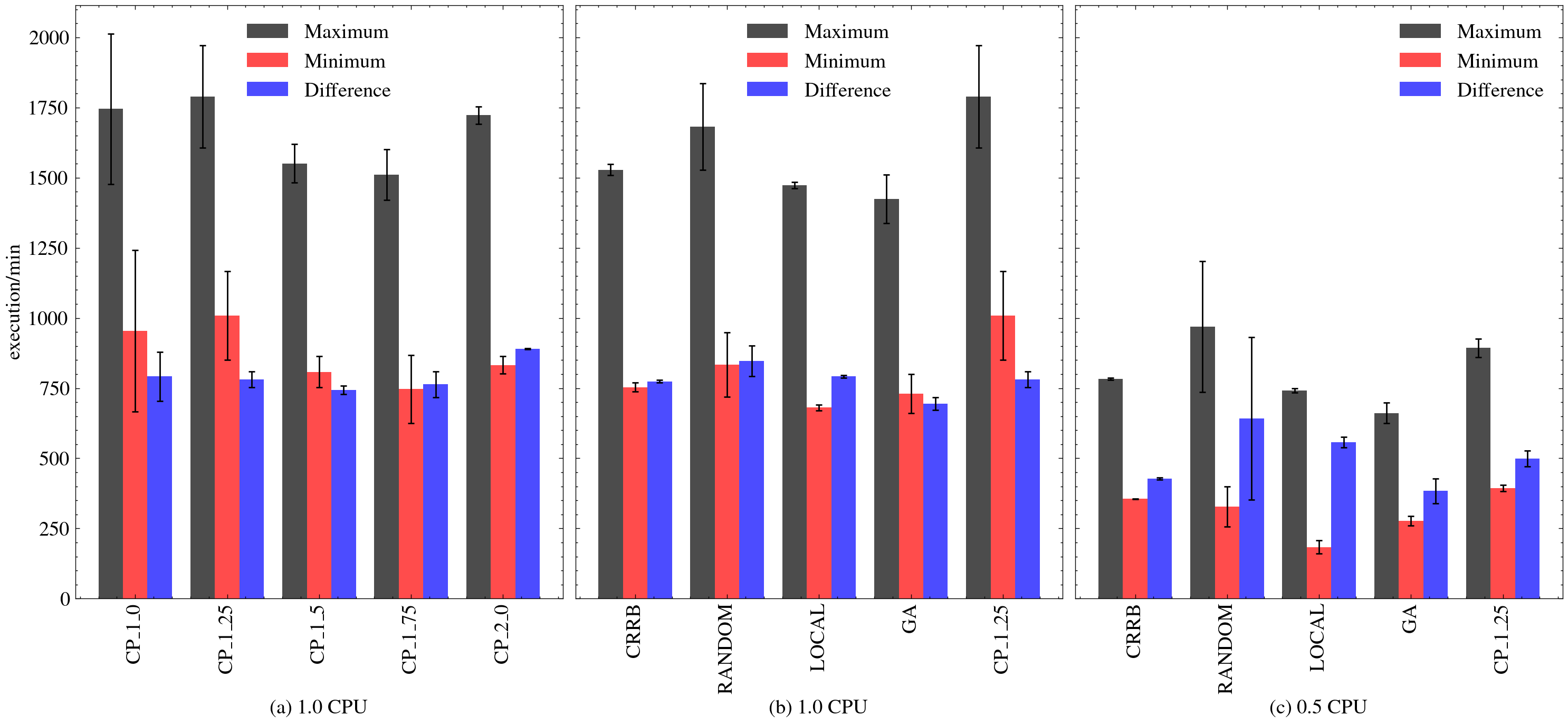}
    \caption{Minimum and maximum event execution count per minute over CEP paths. Leftmost gray bar in each individual distribution result indicates the path with maximum number of event execution, whereas the center red column shows the execution performance of the critical path.}
    \Description{Minimum and maximum event execution count per minute over CEP paths. Leftmost gray bar in each individual distribution result indicates the path with maximum number of event execution, whereas the center red column shows the execution performance of the critical path.}
    \label{min_max_side_by_side}
\end{figure*}

Figure \ref{min_max_side_by_side} shows the maximum and minimum number of sum of events executed per minute on the CEP flow paths. Upon checking the CP only results (a), it can be seen that 1.0 and 1.25 device change penalties show similar results, whereas the 1.0 penalty is more volatile. It should also be noted again that the goal is to increase the performance of the critical path, which is the execution count of the least performing path. Which is the count represented by the "Minimum" value of this figure. Having no penalty (CP\_1\_0) introduces a downside. Optimization disregards the migration cost of the devices and assumes it does not matter which devices are chosen as long as the code and data should be located together or separately are assigned accordingly. On the other side, increasing the penalty forces optimization to keep the assignments as is until not changing it is more costly. Therefore an optimal assignment becomes hard to reach if the initial distribution is not already optimal. Therefore, CP\_1\_25 which is shows the best results and does not ignore the existing assignments is chosen to be compared with heuristics.

Within the same Figure \ref{min_max_side_by_side}, (b) shows that the critical path of CP performs better than the heuristics, and while on average RANDOM performs second best, CRRB and LOCAL have more consistent results as expected due to their deterministic nature. Additionally, GA can sometimes perform better than other heuristic solutions, but on average it performs similarly to CRRB. Lastly, (c) shows the same results under halved CPU power, where CP still performs best. However, this time the mean of RANDOM is less than that of CRRB. It can be seen that sometimes GA performs worse than basic heuristic solutions in terms of having high throughput. One possible reason is that GA requires longer runtime and better individual generation management to reach a solution where it can surpass the heuristics. However, throughout the experiment, reaching an optimal solution not only takes too long in terms of execution time, but also takes several generations across several evaluations. Therefore, the cost of having to migrate code and data due to switching between non-optimal solutions hurts the GA approach. The GA solution can be improved by adding additional mutation and cross functionalities to obtain better individuals in each generation, which is not within the scope of this paper.

\begin{figure*}[ht!]
    \centering
    \includegraphics[width=1.0\linewidth]{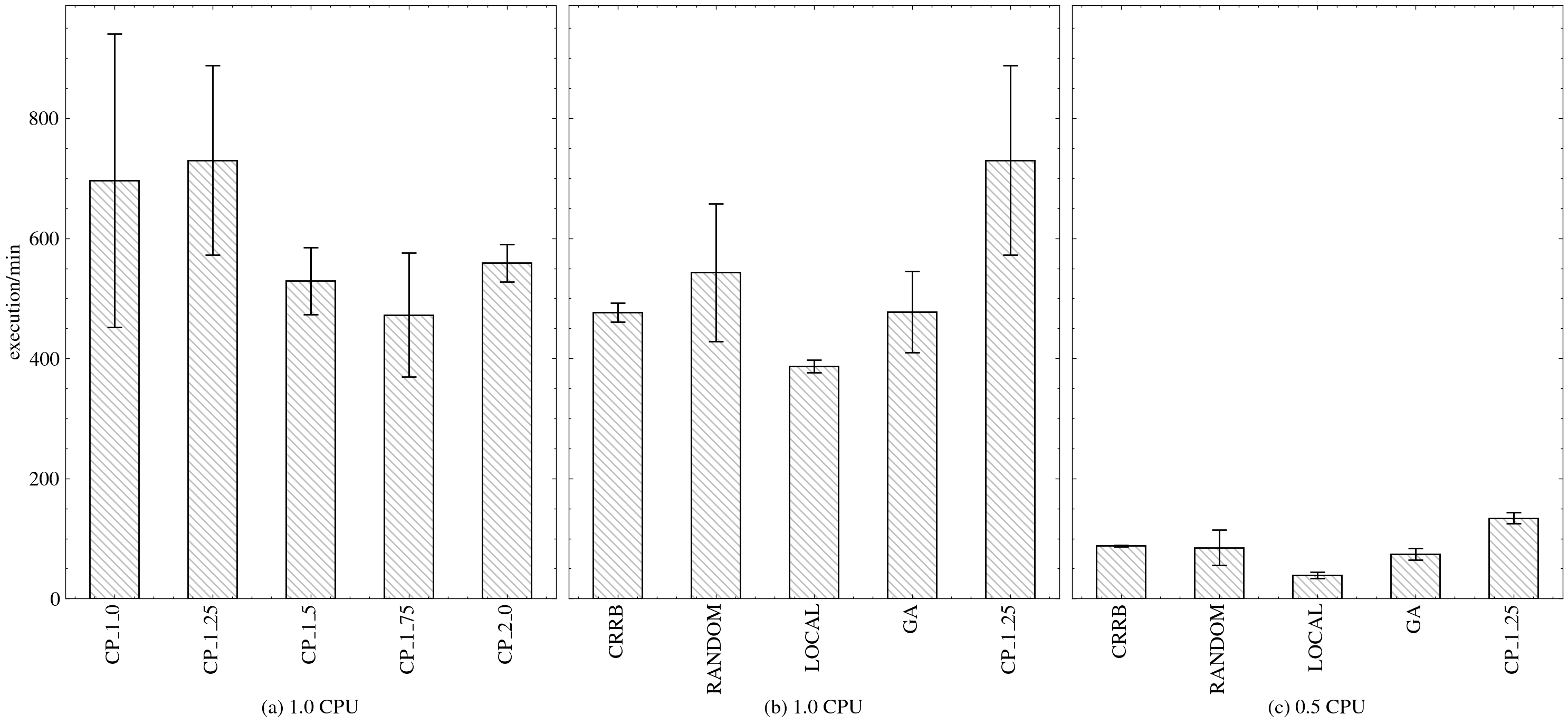}
    \caption{Event execution per minute for the last event of CEP. Last event indicates the event which all paths connect at and therefore is a direct indicator on path performance.}
    \Description{Event execution per minute for the last event of CEP. Last event indicates the event which all paths connect at and therefore is a direct indicator on path performance.}
    \label{last_event_throughput_side_by_side}
\end{figure*}

Figure \ref{last_event_throughput_side_by_side} shows the throughput per minute for the last CEP event only. This figure shows similar results to the previous path Figure \ref{min_max_side_by_side}. This indicates that by increasing the performance of the critical path, the performance of the last event also increases. This holds for all three configurations (a), (b), and (c).

\begin{figure*}[ht]
    \centering
    \includegraphics[width=1.0\linewidth]{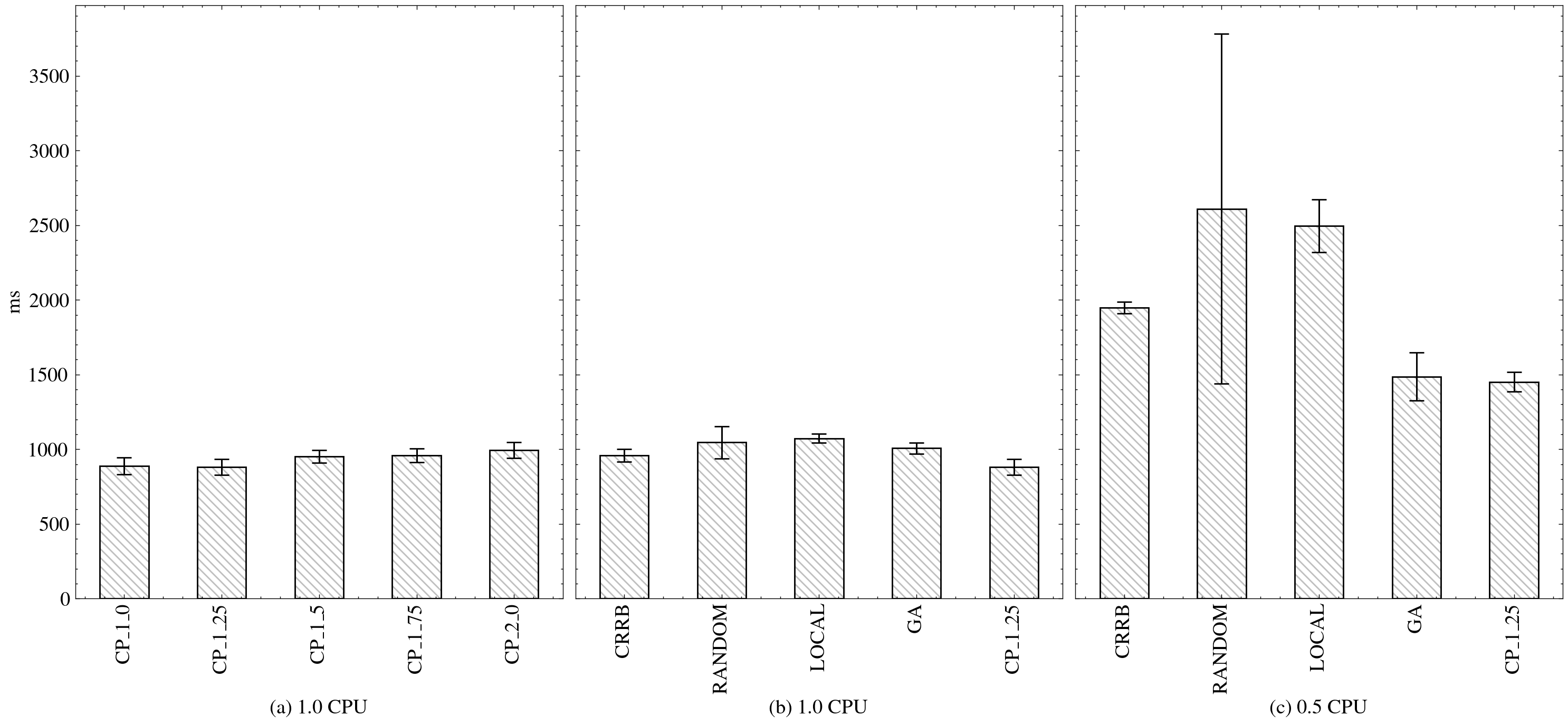}
    \caption{Maximum time it takes to process raw sensor data. An important indicator here is that, within the utilized CEP scenario, having reduced CPU power indicates a huge loss in time.}
    \Description{Maximum time it takes to process raw sensor data. An important indicator here is that, within the utilized CEP scenario, having reduced CPU power indicates a huge loss in time.}
    \label{max_delay_side_by_side}
\end{figure*}

Figure \ref{max_delay_side_by_side} shows how long on average it takes for raw sensor data to be processed throughout the CEP flow paths. Since the experiment tries to replicate a vehicular scenario, these times are significant in terms of the approaches future applicability. The CP results are quite similar, whereas the minimum values belong to CP\_1\_0 and CP\_1\_25. On the other hand, CP again performs better than heuristics, while CRRB is quite close as second best for 1 CPU. and GA as second best for 0.5 CPU. When the CPU power is halved, the difference becomes more obvious, as the performance degradation in GA and CP is much less than that of the other heuristics. An interesting result here is that GA performs well in restricted environments, but ends up having poor throughput. This correlates with the fact that GA requires several generations to reach an optimal result, while it might have reached a solution that has low latency, it can continue to switch between different low latency results, and this constant change damages its throughput due to migration costs.

\begin{figure*}[ht!]
    \centering
    \includegraphics[width=1.0\linewidth]{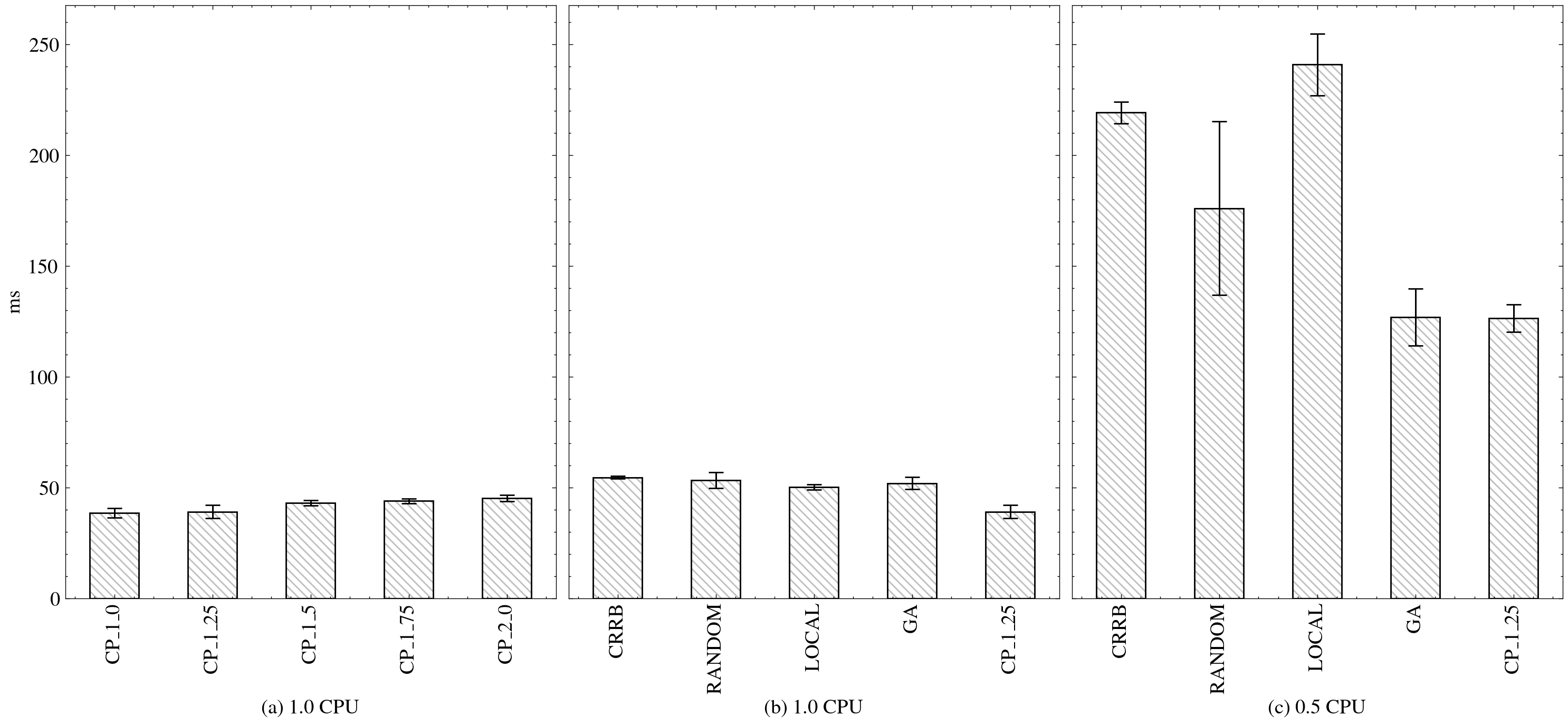}
    \caption{Average execution time for the last event of CEP. Whereas simply checking the last event execution time does not give an indication on path performance, these results show that faster execution exists on the distribution with the better critical path performance.}
    \Description{Average execution time for the last event of CEP. Whereas simply checking the last event execution time does not give an indication on path performance, these results show that faster execution exists on the distribution with the better critical path performance.}
    \label{event_exec_time_side_by_side}
\end{figure*}

\begin{figure*}[ht!]
    \centering
    \includegraphics[width=1.0\linewidth]{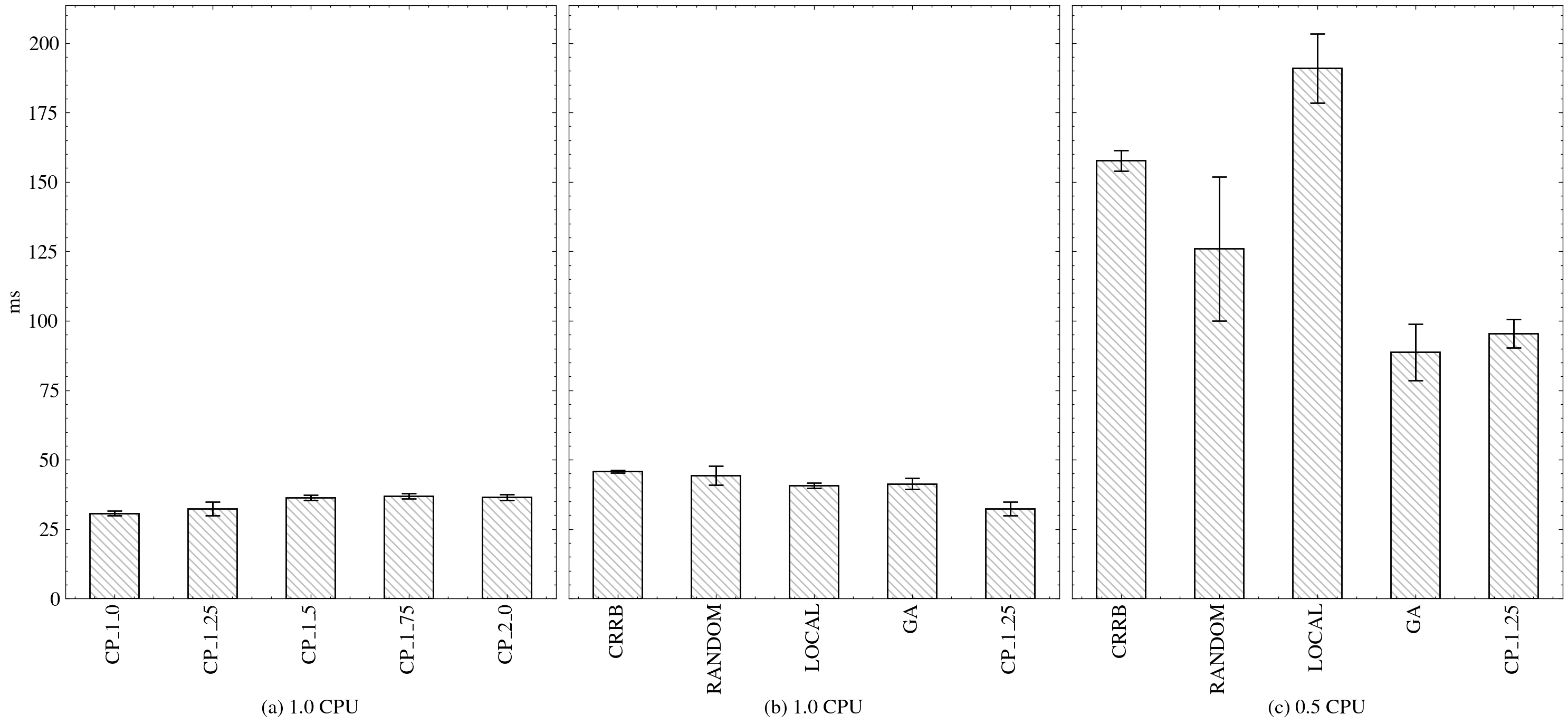}
    \caption{Time it takes to read data for the last event of CEP. Aside from the RANDOM distribution, these results have similar relationship with execution times, which can indicate that data read times may have the biggest impact on execution time.}
    \Description{Time it takes to read data for the last event of CEP. Aside from the RANDOM distribution, these results have similar relationship with execution times, which can indicate that data read times may have the biggest impact on execution time.}
    \label{data_read_times_side_by_Side}
\end{figure*}

Figure \ref{event_exec_time_side_by_side} shows the average execution times for the last CEP event. This time includes the time to read the data, execute the code, and write the result to VSM. The results are quite similar to the throughput as expected. In Figure \ref{data_read_times_side_by_Side} which shows how long it takes to read the input data from VSM for the last CEP event, it can be seen that the data reading times are correlated with the execution times, and it is also the largest factor that contributes to the actual execution time. Therefore, it shows the importance of improving the read times of the data. In these results once again, it can be seen that GA possibly suffers due to the same reason of not converging to an optimal fast enough. Sometimes, it can surpass CP in terms of execution time and data input times but has a higher standard deviation.

\subsection{Scalability} Although the study focuses on small-scale IoT networks, where devices are limited in terms of hardware. A scalability simulation is run to observe how the developed approach performs on different numbers of events and worker devices where each event in a step is tightly connected to all events in the previous step to increase computational complexity. The simulations are run using a single core of the RPi4 devices. When the number of events and workers are both 25, optimization finds a feasible solution around 20 seconds. However, finding an optimal solution takes around 56 seconds. This implies that the optimization may be feasible in non-critical problems up to 25 workers but will be infeasible after that. This correlates with the fact that the number of paths to observe and the number of variables increase faster as the event graph becomes more complex.

\subsection{Limitations}
Doing code migration in the software layer with the download and importing of code files has the advantage of flexibility between different operating systems, as long as the interpreter can be run on the devices. This indicates that the devices should at least have the capability to compile and run custom coding languages. Moreover, since the library is developed using an interpreted language, it can benefit from not having to compile the code files every time they need to be imported. However, this approach can lack in terms of speed compared to cases where the executions were transferred via threads on OS level, which similar approaches were mentioned in the literature section. The management device and the message queues are assumed to work properly, and their existence as a single point of failure has been left out of the context of this study. This study does not address parallelization of the same task across different devices and distributes one task to a single device at any time. Security concerns in IoT devices are not considered within the scope of this study.

\subsection{Threats to Validity} Due to the nature of resource constrained IoT devices, the proposed approach assumes that only a single instance of an event execution should exist at any time. Similarly, a data output should only be written in a single place. This assumption makes the model invalid when parallel executions are required. Additionally, combining data and event executions in optimization may not always be preferred. As it is always possible that some data may not be available to move around due to regulations, privacy, and other reasons. Another reasoning is that whereas combining event and data placement processes together has the benefit of also including the dependencies, it increases the optimization time which in cases where event stream load is high, the optimization may negatively affect the QoL requirements. That is why the study focuses on restricted devices with certain load capabilities.

\section{Conclusion}
Complex event processing solutions in IoT environments need to address the limited capabilities of the devices to prevent possible performance issues when multiple devices need to work synchronously. An increased complexity of tasks required to run on these devices requires better execution and data management. In this study, the challenges of using dependent CEP tasks in the IoT environment are explored. We have applied optimization over the critical path of the task dependency graph and implemented a library to address performance issues when code and data location assignments are sub-optimal. The library enables devices to access each others' event data through a virtually shared memory layer. Furthermore, custom code executions that enable more complex tasks in the event flow are included in the optimization. Multiple distribution algorithms are compared with the new approach. The results of experimenting on a real-world scenario indicate that by improving critical path performance, it is possible to improve overall throughput and I/O delay.

Future work and possible research areas can be summarized as follows. Worker execution data can be utilized with ML, discrete optimization, and graph algorithms to faster adapt the CEP flow. Optimization in more complex CEP flows can be studied and the effects of the proposed study on energy consumption can be explored. The problem of scalability in smart cities can be explored, as multiple same actions can be executed across different devices for cases such as city-wide alarms. In such a scenario, data partitioning for faster access, regional DAG algorithms and sub-graph creation for reducing graph complexity, multiple management devices to reduce traffic overload can help improve the performance of the overall CEP requirement. The loss of critical short-term data on hardware and software failure can be explored to increase the overall resiliency of the infrastructure.

\section*{Acknowledgments} This work is jointly supported by The Scientific and Technological Research Council of Turkey (TUBITAK) 1515 Frontier R\&D Laboratories Support Program for BTS Advanced AI Hub: BTS Autonomous Networks and Data Innovation Lab. Project 5239903 and Research Fund of the Istanbul Technical University, Project MCAP-2022-43825.

\bibliographystyle{unsrt}
\bibliography{bibliography}

\end{document}